\documentstyle[prl,aps,floats,psfig]{revtex}
\begin{document}
\draft
\preprint{gr-qc/9911046}
\title{Regular Black Hole in General Relativity Coupled to Nonlinear Electrodynamics}
\author{Eloy Ay\'on--Beato\thanks{%
On leave from Grupo de F\'\i sica Te\'orica, CEMAFIT--ICIMAF, Calle E $\#$
309, esq.~a 15, CP 10400, Ciudad Habana, Cuba.} and Alberto Garc\'\i a}
\address{Departamento~de~F\'{\i}sica,~Centro~de~Investigaci\'on~%
y~Estudios~Avanzados~del~IPN\\
Apdo. Postal 14--740, 07000 M\'exico DF, MEXICO}
\maketitle

\begin{abstract}
The first regular {\em exact\/} black hole solution in General Relativity is
presented. The source is a nonlinear electrodynamic field satisfying the
weak energy condition, which in the limit of weak field becomes the Maxwell
field. The solution corresponds to a charged black hole with $|q|\leq 2s_{%
{\rm {c}}}m\approx 0.6\,m$, having the metric, the curvature invariants, and
the electric field regular everywhere.
\end{abstract}


In General Relativity the existence of singularities appears to be a
property inherent to most of the physically relevant solutions of Einstein
equations, in particular, to all known up--to--date black hole {\em exact\/}
solutions \cite{H-E}. The Penrose cosmic censorship conjecture states that
these singularities must be dressed by event horizons; no causal connection
could exist between the interior of a black hole with the exterior fields,
thus pathologies occurring at the singular region would have no influence on
the exterior region, and the Physics outside would be well behaved (cf.~\cite
{Wald97} for a review on the recent status of this conjecture).

To avoid the black hole singularity problem, some regular {\em models} has
been proposed \cite
{Bardeen68,Ayon93,Borde94,BarrFrolov96,MMPSenovilla96,CaboAyon97}. All of
them have been referred to as ``Bardeen black holes'' \cite{Borde97}, since
Bardeen was the first author producing a surprising regular black hole model 
\cite{Bardeen68}. No one of these models is an {\em exact\/} solution to
Einstein equations; there are no known physical sources associated with any
of them. The attempts to solve this problem have usually been addressed to
the search of more general gravity theories. The best candidate today to
produce singularity--free solutions, even at the classical level, due to its
intrinsic non--locality, is string theory \cite{Tseytlin95}. There are
examples in other contexts, for instance, in $N=1$ supergravity domain wall
solutions with horizons but no singularities have been found (cf.~\cite
{Cvetic93}, and references therein), another example is given in exact
conformal field theory \cite{HHorowitz92}.

We show in this Letter that in the framework of the standard General
Relativity one can find singularity--free solutions of the Einstein field
equations coupled to a suitable nonlinear electrodynamics, which in the weak
field approximation becomes the usual linear Maxwell theory. Previous
efforts on this direction with nonlinear electrodynamics either have been
totally unsuccessful or only partially solve the considered singularity
problem \cite{Oliveira94,Soleng95,Palatnik97}. We propose a new nonlinear
electrodynamics which coupled to gravity actually produces a non--singular 
{\em exact\/} black hole solution satisfying the weak energy condition.

\begin{figure}[h]
\centerline{ \psfig{file=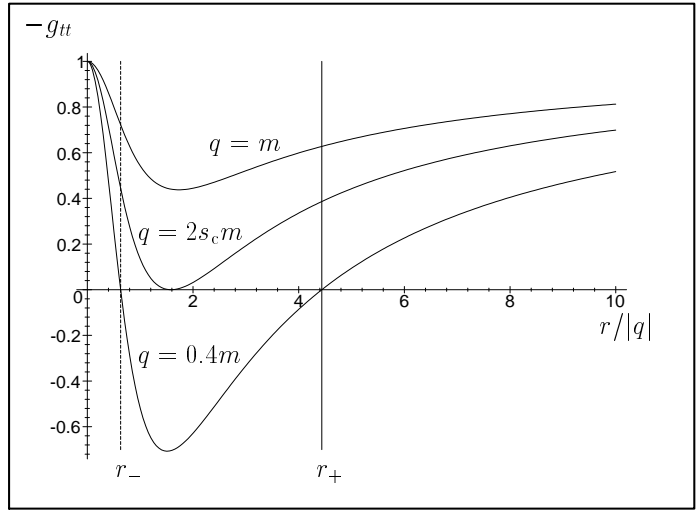,width=10cm} }
\caption{Behavior of $-g_{tt}$ for different values of charge.}
\label{fig:A}
\end{figure}

\begin{figure}[h]
\centerline{\ \psfig{file=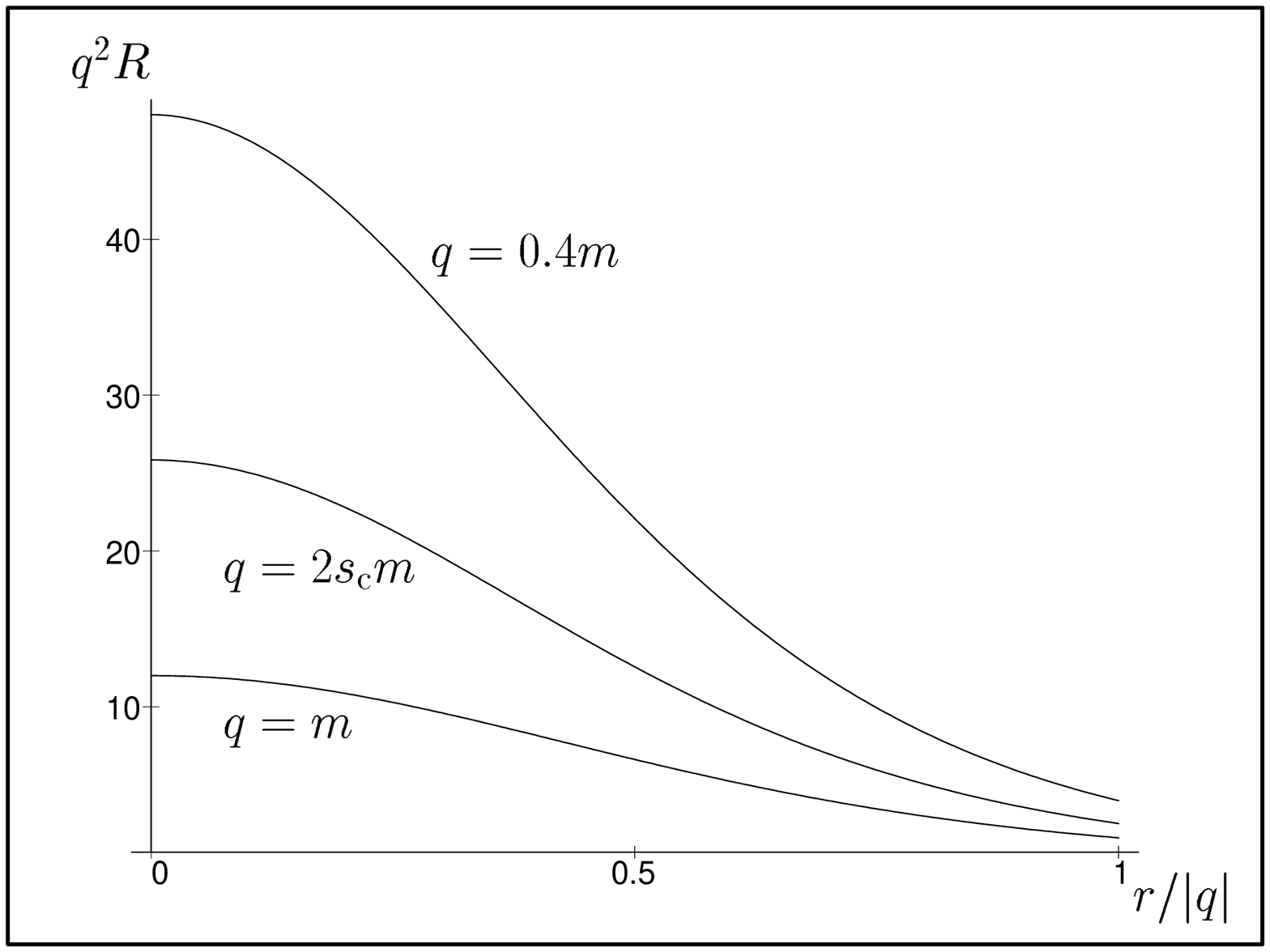,width=6.8cm} %
\psfig{file=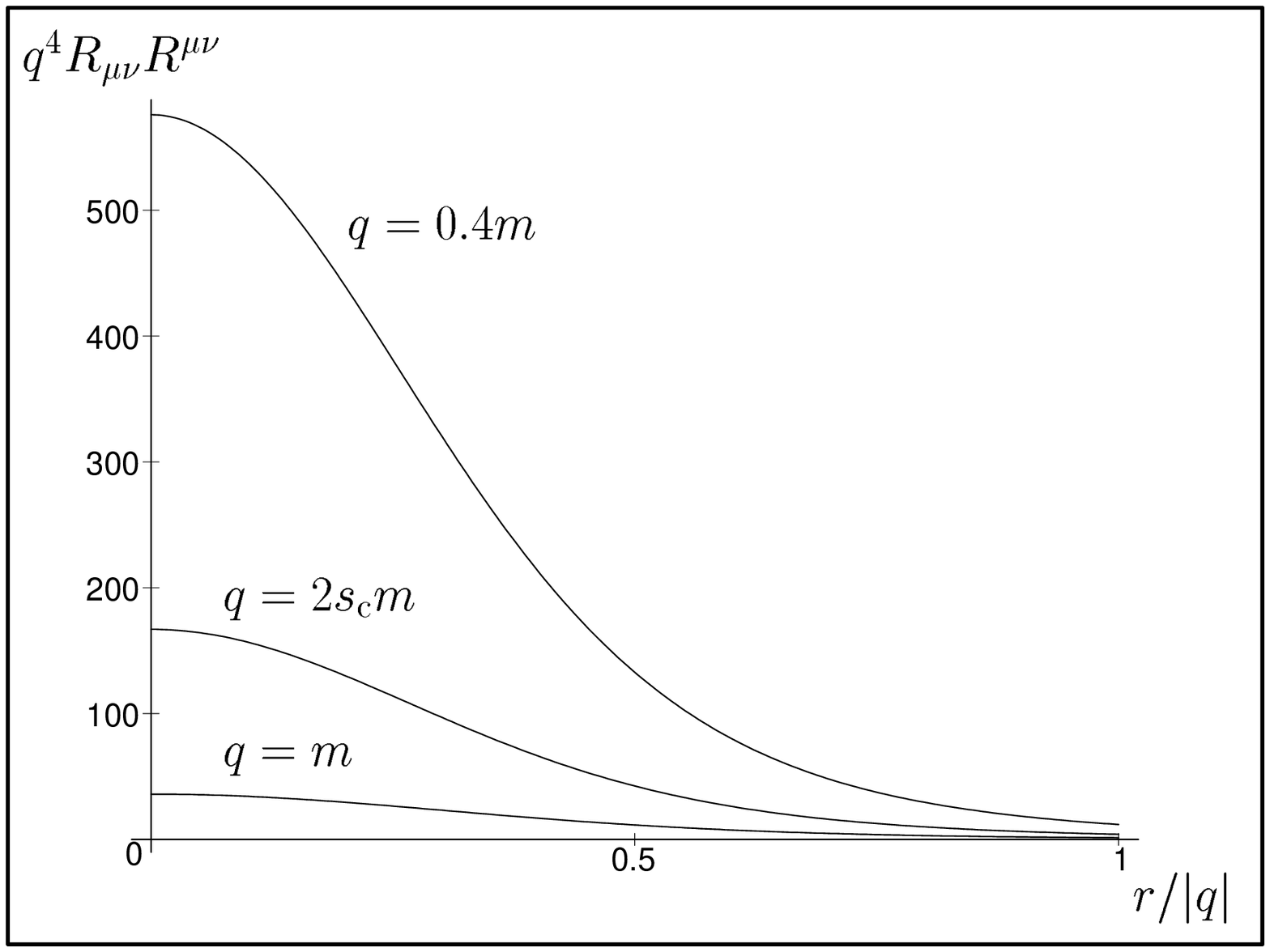,width=6.8cm} %
\psfig{file=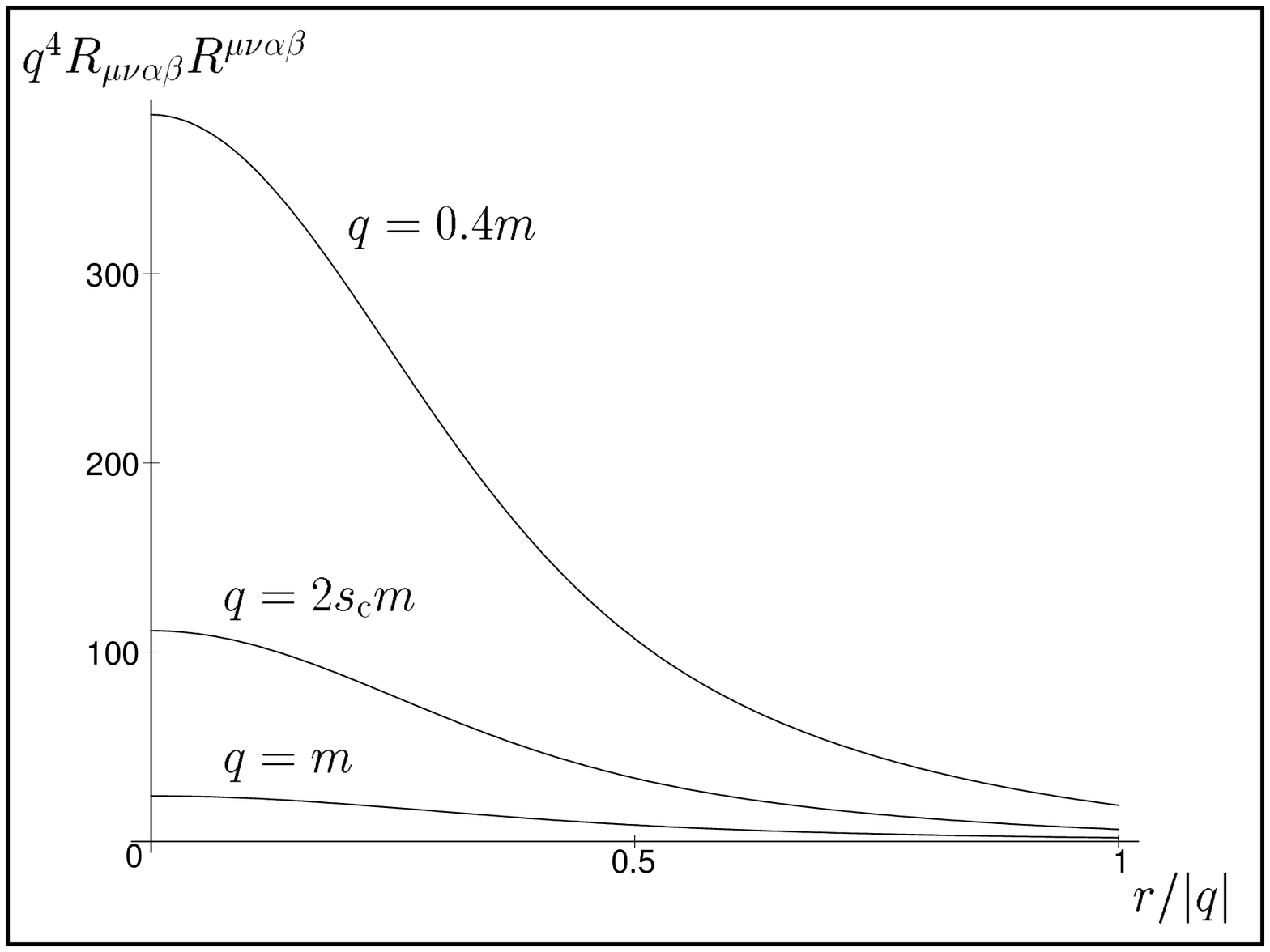,width=6.8cm} }
\caption{Regular behavior of the Ricci, $q^2R$, Ricci square, $q^4R_{\mu \nu
}R^{\mu \nu }$, and the Riemann square, $q^4R_{\mu \nu \alpha \beta }R^{\mu
\nu \alpha \beta }$, scalars for different values of charge; the abscissa is 
$r/|q|$.}
\label{fig:curv}
\end{figure}

The gravitational field of our solution is described by the metric 
\begin{equation}
\text{\boldmath$g$}=-\left( 1-\frac{2mr^2}{(r^2+q^2)^{3/2}}+\frac{q^2r^2}{%
(r^2+q^2)^2}\right) \text{\boldmath$dt$}^2+\left( 1-\frac{2mr^2}{%
(r^2+q^2)^{3/2}}+\frac{q^2r^2}{(r^2+q^2)^2}\right) ^{-1}\text{\boldmath$dr$}%
^2+r^2\text{\boldmath$d\Omega $}^2,  \label{eq:regbh}
\end{equation}
while the associated electric field $E$ is given by 
\begin{equation}
E=q\,r^4\left( \frac{r^2-5\,q^2}{(r^2+q^2)^4}+\frac{15}2\,\frac m{%
(r^2+q^2)^{7/2}}\right) .  \label{eq:E}
\end{equation}
Notice that this solution asymptotically behaves as the
Reissner--Nordstr\"om solution, i.e., 
\[
-g_{tt}=1-2m/r+q^2/r^2+O(1/r^3),\qquad E=q/r^2+O(1/r^3),
\]
thus the parameters $m$ and $q$ are related correspondingly with the mass
and the electric charge. For a certain range of the mass and charge our
metric (\ref{eq:regbh}) is a black hole, which in addition is regular
everywhere. Accomplishing the substitutions $x=r/|q|$ and $s=|q|/2m$, we
rewrite $g_{tt}$ as 
\begin{equation}
-g_{tt}=A(x,s)\equiv 1-\frac 1s\frac{x^2}{(1+x^2)^{3/2}}+\frac{x^2}{(1+x^2)^2%
},  \label{eq:A}
\end{equation}
which, for any nonvanishing value of $s$, has a single minimum; cf.~Fig.~\ref
{fig:A}. 
There exists a single real critical value of $x$, $x_{{\rm c}}$, and one of $%
s$, $s_{{\rm c}}$, to be determined from $A(x_{{\rm {c}}},s_{{\rm {c}}})=0$
and $\partial _xA(x_{{\rm {c}}},s_{{\rm {c}}})=0$, namely 
\[
t^4-\frac{t^3}s+t^2+\frac ts-1=0,\qquad \frac{t^3}s-2t^2-\frac{3t}s+4=0,
\]
where $t^2\equiv {x^2+1}$. To solve these equations, one substitutes $%
s=t(t^2-3)/(2t^2-4)$ from the second equation into the first one arriving at 
$t^6-4t^4+2t^2-1=0$, which has only one real solution for $t^2$, thus the
corresponding critical values are $s_{{\rm {c}}}\approx 0.317$ and $x_{{\rm {%
c}}}\approx 1.58$. For $s<s_{{\rm {c}}}$ the quoted minimum is negative, for 
$s=s_{{\rm {c}}}$ the minimum vanishes, and for $s>s_{{\rm {c}}}$ the
minimum is positive. Evaluating the curvature invariants $R$, $R_{\mu \nu
}R^{\mu \nu }$, and $R_{\mu \nu \alpha \beta }R^{\mu \nu \alpha \beta }$ for
metric (\ref{eq:regbh}) one establishes that they are all regular
everywhere, cf.~Fig.~\ref{fig:curv}; thus for $s\leq s_{{\rm {c}}}$ the
singularities appearing in (\ref{eq:regbh}) due to the vanishing of $A$ are
only coordinate singularities describing the existence of horizons,
consequently, we are in the presence of black hole solutions for $|q|\leq
2s_{{\rm {c}}}m\approx 0.6\,m$. 
For these values of mass and charge we have, under the strict inequality $%
|q|<2s_{{\rm {c}}}m$, inner and event horizons for the Killing field $\text{%
\boldmath$k$}=\text{\boldmath$\partial /\partial {t}$}$, defined by the real
solutions of the quartic equation $-k_\mu k^\mu =A=0$, which are given by 
\begin{equation}
r_{\pm }=|q|\left( \left[ \frac 1{4s}+\frac{\sqrt{f(s)}}{12s}\pm \frac{\sqrt{%
6}}{12s}\left( \frac 92-12s^2-\frac{f(s)}6-\frac{9(12s^2-1)}{\sqrt{f(s)}}%
\right) ^{1/2}\right] ^2-1\right) ^{1/2},  \label{eq:horiz}
\end{equation}
\[
f(s)=6\left( \frac 32-4\,s^2+s\,g(s)^{1/3}-\frac{4s\left( 11\,s^2-3\right) }{%
\,g(s)^{1/3}}\right) ,
\]
\[
g(s)=4\left( 9\,s+74\,s^3+\sqrt{27(400\,s^6-112\,s^4+47\,s^2-4)}\right) .
\]
For $|q|=2s_{{\rm {c}}}m$, the horizons shrink into a single one,
corresponding to an extreme black hole ($\nabla _\nu (k_\mu k^\mu )=0$). The
extension of the metric beyond the horizons $r_{\pm }$ becomes apparent by
passing to the standard advanced and retarded Eddington--Finkelstein
coordinates, in terms of which the metric is smooth everywhere, even in the
extreme case. Following step by step the procedure presented in \cite[Chap.V]
{Chandra83} to derive the global structure of the Reissner--Nordstr\"om
black hole, one can arrive at the global structure of our solution and
construct the Penrose diagrams; nevertheless, because of journal length
restrictions we omit here the corresponding calculations and diagrams,
leaving this issue for an extended publication. Briefly, what one encounters
in the case of our non--extreme black hole solution, $|q|<2s_{{\rm {c}}}m$,
is the splitting of the space--time into three regions, I: $r>r_{+}$, II: $%
r_{-}<r<r_{+}$, and III: $0\leq r<r_{-}$; cf.~Fig.~\ref{fig:A}. In each
region one introduces advanced and retarded coordinates $u$ and $v$, related
with $r$ through the so called tortoise coordinate $r^{*}\equiv \int A^{-1}dr
$, which in our case is quite involved. Further, by the inversion of $u$ and 
$v$, $u\rightarrow -u$, $v\rightarrow -v$, one obtains the remaining regions
I$^{^{\prime }}$, II$^{^{\prime }}$, and III$^{^{\prime }}$. Introducing a
new set of null coordinates one arrives at the maximal extension of the
non--extreme black hole. The Penrose diagram of the maximal analytical
extension of our solution is obtained by gluing appropriately copies of
these six regions upward and downward {\em ad infinitum}. In the extreme
black hole case, $|q|=2s_{{\rm {c}}}m$, there arise two regions, I: $r>r_{%
{\rm {c}}}$ and III: $0\leq r<r_{{\rm {c}}}$, cf.~Fig.~\ref{fig:A}, in which
again one introduces advanced and retarded $u$ and $v$ coordinates to
accomplish the maximal analytical extension; these two region determine the
main building block of the extension. To construct the Penrose diagram of
the maximal analytical extension, one glues copies of this block in a
suitable way. In both cases, extreme and non--extreme, there is no
singularity at $r=0$, which is now simply the origin of the spherical
coordinates. Summarizing, our space--time possesses the same global
structure as the Reissner--Nordstr\"om black hole except that the
singularity, at $r=0$, of this last solution has been smoothed out.

For $|q|>2s_{{\rm {c}}}m$, there are no horizons and the corresponding exact
solution represents a globally regular space--time. It is worthwhile to
mention in this respect the existence of globally smooth solutions to the
Einstein+matter (Yang--Mills, Yang--Mills--Higgs) equations; although there
are demonstrations of the existence of these solutions \cite
{Smoller93,Breiten94}, they are numerically given and there are no
analytical closed expressions for them \cite{BartnikMcKi88}; cf.~\cite
{Bizon94}, and references therein.

\begin{figure}[h]
\centerline{\ \psfig{file=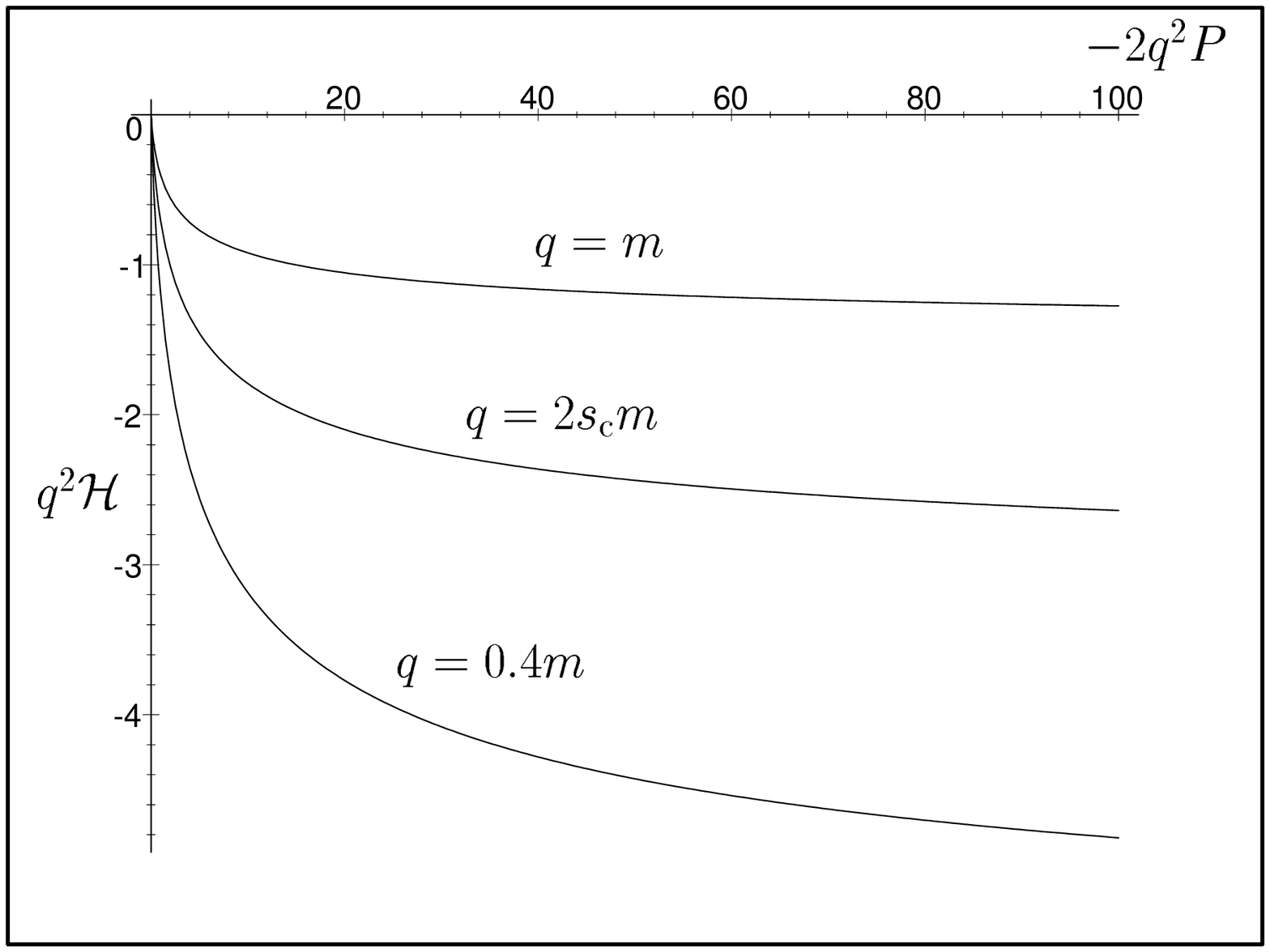,width=10cm} %
\psfig{file=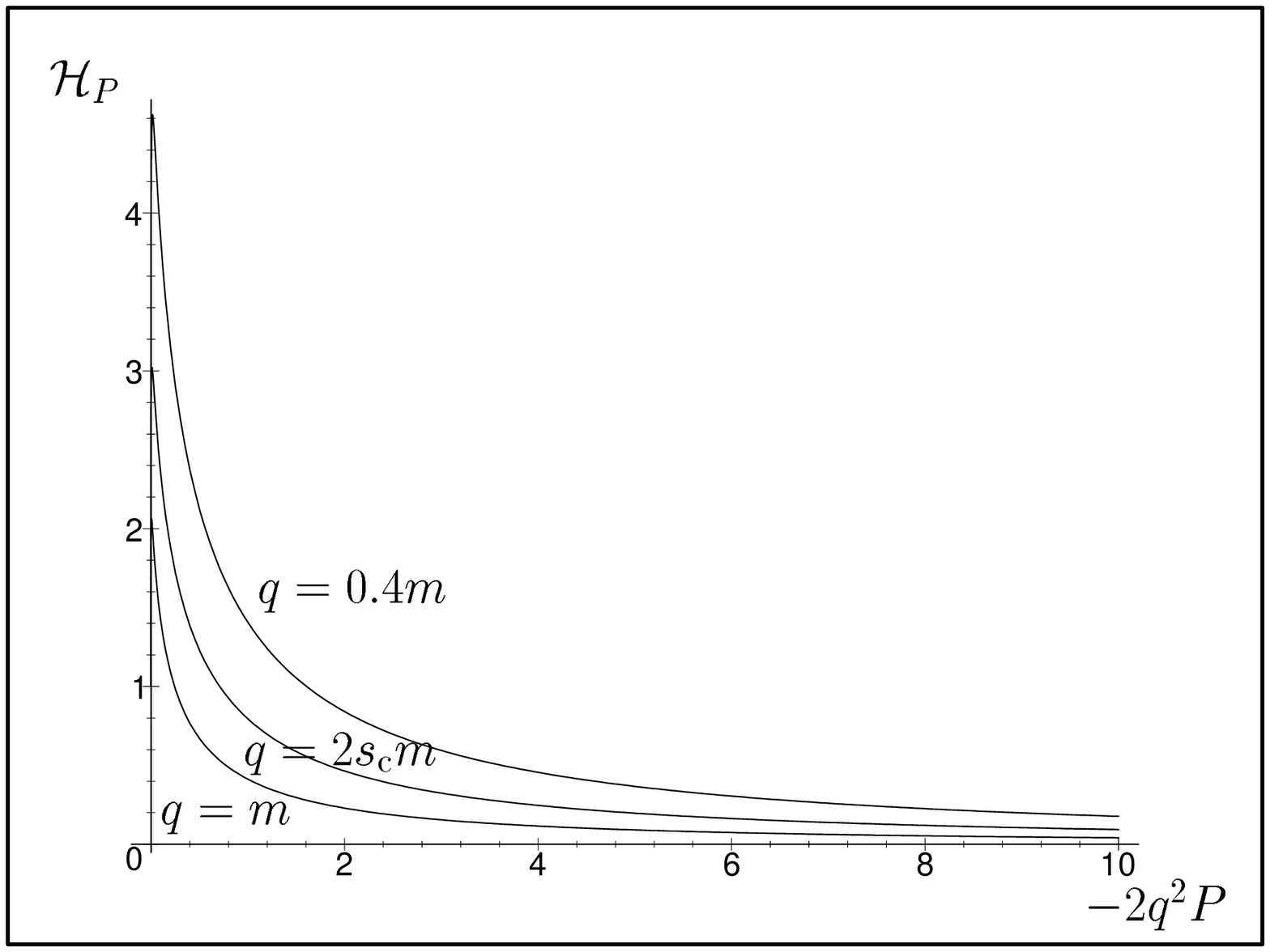,width=10cm} }
\caption{Behavior of $q^2{\cal H}$ and ${\cal H}_P$ with respect to the
positive abscissa $-2q^2P$ for different values of charge.}
\label{fig:wec}
\end{figure}

The fields (\ref{eq:regbh}) and (\ref{eq:E}) arise as a solution of the
Einstein--nonlinear electrodynamic field equations derived from the action
proposed in Einstein--dual nonlinear electrodynamic theory \cite{SGP87},
which in the studied case becomes 
\begin{equation}
{\cal S}=\int dv\left( \frac 1{16\pi }R-\frac 1{4\pi }{\cal L}(F)\right) ,
\label{eq:action}
\end{equation}
where $R$ is scalar curvature, and ${\cal L}$ is a function of $F\equiv 
\frac 14F_{\mu \nu }F^{\mu \nu }$. Alternatively, one can describe the
considered system using another function obtained by means of a Legendre
transformation \cite{SGP87}: 
\begin{equation}
{\cal H}\equiv 2F{\cal L}_F-{\cal L}.  \label{eq:Leg}
\end{equation}
Defining $P_{\mu \nu }\equiv {\cal L}_FF_{\mu \nu }$, it can be shown that $%
{\cal H}$ is a function of $P\equiv \frac 14P_{\mu \nu }P^{\mu \nu }=({\cal L%
}_F)^2F$, i.e., $d{\cal H}=({\cal L}_F)^{-1}d(({\cal L}_F)^2F)={\cal %
H}_PdP$. With the help of ${\cal H}$ one expresses the nonlinear
electromagnetic Lagrangian in the action (\ref{eq:action}) as ${\cal L}=2P%
{\cal H}_P-{\cal H}$, depending on the anti--symmetric tensor $P_{\mu \nu }$%
. The specific function ${\cal H}$, determining the nonlinear electrodynamic
source used, is given as 
\begin{equation}
{\cal H}(P)=P\,\frac{\left( 1-3\sqrt{-2\,q^2P}\right) }{\left( 1+\sqrt{%
-2\,q^2P}\right) ^3}-\frac 3{2\,q^2s}\left( \frac{\sqrt{-2\,q^2P}}{1+\sqrt{%
-2\,q^2P}}\right) ^{5/2},  \label{eq:H}
\end{equation}
where $s=|q|/2m$ and the invariant $P$ is a negative quantity. The
corresponding Lagrangian occurs to be 
\begin{equation}
{\cal L}=P\,\frac{\left( 1-8\sqrt{-2\,q^2P}-6\,q^2P\right) }{\left( 1+\sqrt{%
-2\,q^2P}\right) ^4}-\frac 3{4\,q^2s}\frac{(-2\,q^2P)^{5/4}\left( 3-2\sqrt{%
-2\,q^2P}\right) }{\left( 1+\sqrt{-2\,q^2P}\right) ^{7/2}}.  \label{eq:Lagex}
\end{equation}
The function (\ref{eq:H}) satisfies the plausible conditions, needed for a
nonlinear electromagnetic model, of (i) correspondence to Maxwell theory, 
i.e., ${\cal H}\approx P$ for weak fields ($P\ll 1$), and (ii) the
weak energy condition, which requires ${\cal H}<0$ and ${\cal H}_P>0$;
cf.~Fig.~\ref{fig:wec}. We would like to point out that our solution, in
addition to being regular and to satisfying the weak energy condition, is
characterized by another feature: it does not admit a Cauchy surface. Hence,
it does not contradict the Penrose singularity theorem supported on the
hypotheses of: fulfillment of the null energy condition, existence of a
noncompact Cauchy surface, and existence of a closed trapped surface and
concluding no null geodesically completeness of the space--time. 

In what follows we shall briefly give the main lines of the integration
process yielding the studied solution. The Einstein and nonlinear
electrodynamic equations arising from action (\ref{eq:action}) are 
\begin{equation}
G_\mu ^{~\nu }=2({\cal H}_PP_{\mu \lambda }P^{\nu \lambda }-\delta _\mu
^{~\nu }(2P{\cal H}_P-{\cal H})),  \label{eq:Ein}
\end{equation}
\begin{equation}
\nabla _\mu P^{\alpha \mu }=0.  \label{eq:Max}
\end{equation}
In order to obtain the solution (\ref{eq:regbh}), (\ref{eq:E}), we consider
the static and spherically symmetric configuration 
\begin{equation}
\text{\boldmath$g$}=-\left( 1-\frac{2m}r+\frac{Q(r)}{r^2}\right) \text{%
\boldmath$dt$}^2+\left( 1-\frac{2m}r+\frac{Q(r)}{r^2}\right) ^{-1}\text{%
\boldmath$dr$}^2+r^2\text{\boldmath$d\Omega $}^2,  \label{eq:spher}
\end{equation}
and the following ansatz for the antisymmetric field $P_{\mu \nu }=2\delta
_{[\mu }^t\delta _{\nu ]}^rD(r)$. With these choices the equations (\ref
{eq:Max}) integrate as 
\begin{equation}
P_{\mu \nu }=2\delta _{[\mu }^t\delta _{\nu ]}^r\frac q{r^2}\quad
\longrightarrow \quad P=-\frac{D^2}2=-\frac{q^2}{2r^4},  \label{eq:dielec}
\end{equation}
where we have chosen the integration constant as $q$ since, as it was
previously anticipated, it actually plays the role of the electric charge.
The evaluation of the electric field $E=F_{tr}={\cal H}_PD$, using
expression (\ref{eq:H}) for ${\cal H}$, gives just the formula (\ref{eq:E}).
The $_t^{~t}$ component of Einstein equations (\ref{eq:Ein}) yields the
basic equation 
\begin{equation}
\frac{rQ^{\prime }-Q}{r^4}=2{\cal H}(P).  \label{eq:tt}
\end{equation}
Substituting ${\cal H}$ from (\ref{eq:H}) with $P=-q^2/2r^4$ one can write
the integral of (\ref{eq:tt}) as 
\begin{equation}
Q=q^2r\int_r^\infty dy\left( \frac{6my^2}{(y^2+q^2)^{5/2}}+\frac{%
y^2(y^2-3q^2)}{(y^2+q^2)^3}\right) ,  \label{eq:int}
\end{equation}
the integrand above can be expressed as $\partial
_y(2my^3/q^2(y^2+q^2)^{3/2}-y^3/(y^2+q^2)^2)$, thus one arrives at 
\begin{equation}
Q=2mr-\frac{2mr^4}{(r^2+q^2)^{3/2}}+\frac{q^2r^4}{(r^2+q^2)^2}.  \label{eq:Q}
\end{equation}
Substituting $Q$ into $-g_{tt}=1-2m/r+Q/r^2$ one finally gets Eq. (\ref
{eq:regbh}).

\acknowledgments
This work was partially supported by the CONACyT Grant 3692P--E9607, and a
fellowship from the Sistema Nacional de Investigadores (SNI). One of the
authors E.A.B. thanks the staff of the Physics Department at CINVESTAV for
support.

\end{document}